\documentclass[final,1p,times]{elsarticle}

\usepackage{aas_macros}

\usepackage[T1]{fontenc}
\usepackage[utf8]{inputenc}
\usepackage[english]{babel}
\usepackage{amsmath}
\usepackage{amssymb}
\usepackage{bm}

\usepackage{graphicx}
\usepackage{subfig}
\usepackage{enumerate}
\usepackage{epsfig,epstopdf}
\usepackage{color,xcolor}
\usepackage{multirow}
\usepackage{mleftright}
\usepackage{soul}
\usepackage{float}

\usepackage{url}
\usepackage{xspace}
\usepackage{comment}
\usepackage{braket}
\usepackage{hyperref}
\usepackage{cleveref}
\usepackage{hhline}

\biboptions{sort&compress}

\newcommand{\veps}{\varepsilon}
\newcommand{\gCDM}{$\gamma {\rm CDM}\,$}
\newcommand{\wCDM}{$w_0w_a{\rm CDM}\,$}
\newcommand{\bk}{\bm{k}}

\newcommand{\bx}{\bm{x}}

\long\def\comment#1{}

\newcommand*\mean[1]{\overline{#1}}

\newcommand{\Planck}{\textit{Planck}\,}

\graphicspath{{./plots/}}

\journal{Physics of the Dark Universe}

\begin{document}

\begin{frontmatter}

\title{{Cosmological constraints from the EFT power spectrum and tree-level bispectrum of 21cm intensity maps}}

\author[1]{Liantsoa F. Randrianjanahary\corref{cor1}}
\ead{fina.liantsoarandrianjanahary@gmail.com}
\cortext[cor1]{Corresponding author.}
\author[1]{Dionysios Karagiannis}
\author[1,2,3]{Roy Maartens}

\affiliation[1]{Department of Physics $\&$ Astronomy, University of the Western Cape, Cape Town 7535, South Africa}
\affiliation[2]{National Institute for Theoretical $\&$ Computational Sciences (NITheCS), Cape Town 7535, South Africa}
\affiliation[3]{Institute of Cosmology $\&$ Gravitation, University of Portsmouth, Portsmouth PO1 3FX, United Kingdom}

\begin{abstract}
We explore the information content of 21cm intensity maps in redshift space using the 1-loop Effective Field Theory power spectrum model and the bispectrum at tree level. The 21cm signal contains signatures of dark matter, dark energy and the growth of large-scale structure in the Universe. These signatures are typically analyzed via the 2-point correlation function or power spectrum. However, adding the information from the 3-point correlation function or bispectrum will be crucial to exploiting next-generation intensity mapping experiments. The bispectrum could offer a unique opportunity to break key parameter degeneracies that hinder the measurement of cosmological parameters and improve on the precision. We use a Fisher forecast analysis to estimate the constraining power of the HIRAX survey on cosmological parameters, dark energy and modified gravity.
\end{abstract}


\end{frontmatter}

\section{Introduction}

The effective field theory of large-scale structure (EFTofLSS) is a formalism developed to accurately and reliably predict the clustering of cosmological large-scale structure in the mildly nonlinear regime \cite{Senatore:2014vja,Angulo_2015,Foreman_2016,d_Amico_2020,Ivanov:2022mrd}.
EFTofLSS differs significantly from the widely used standard perturbation theory (SPT) formalism  \cite{Bernardeau2002}. The fundamental difference arises from the fact that EFTofLSS considers the impact of nonlinearities on scales that are only mildly nonlinear by introducing effective stresses into the equations of motion. As a result, the 1-loop power spectrum is modified by adding ultra-violet counterterms that account for the influence of short-distance physics on long-distance scales. 

EFTofLSS provides a means to perform analytical computations of LSS observables with the required precision in the mildly nonlinear regime. \cite{Cataneo:2016suz} develop efficient implementations of these computations that explore their dependence on cosmological parameters. \cite{Naskar:2017ekm} investigate the possibility of constraining parameters of the EFT of inflation with upcoming LSS surveys. \cite{Ivanov:2020ril} perform the first early dark energy analysis using the full-shape power spectrum likelihood from the Baryon Oscillation Spectroscopic Survey (BOSS), based on EFTofLSS. \cite{d_Amico_2020} analyses the cosmological parameters in the DR12 BOSS dataset after validating the approach with various numerical simulations.

EFTofLSS requires several new bias coefficients 
that contribute to multiple observables. A set of seven bias parameters adequately characterizes the manifold statistical properties \cite{Angulo_2015_bis,perko2016biased,Fujita_2020}.
EFTofLSS has provided a robust framework to describe large-scale galaxy distributions without requiring detailed knowledge of galaxy formation physics. It overcomes the limitations of the pressureless perfect fluid representation by treating dark matter as an effective non-ideal fluid. This theory is based on time-sliced perturbation theory, a non-equilibrium field theory approach that allows consistent renormalization of cosmological correlation functions and systematic resummation of large infrared effects.
The equations of motion within the EFTofLSS are derived through a perturbative series expansion of matter fields in density contrast. The expansion parameter is the ratio of the scale of interest to the nonlinear scale of matter distribution. 
EFTofLSS strategically uses only the relevant degrees of freedom for large-scale dynamics, effectively parametrizing the impact of unknown short-scale physics. The framework has undergone extensive validation against data \cite{Zhang_2022,Carrilho_2023}, demonstrating excellent agreement on mildly-nonlinear scales.

The EFTofLSS equations of motion form a set of nonlinear and non-local partial differential equations, effectively describing the evolution of density and velocity fields on large scales. This versatile framework facilitates the prediction of large-scale structure in the Universe and enables the study of various physical processes, such as baryons, neutrinos, and dark energy.
The EFTofLSS framework has become useful for figuring out the mysteries of large-scale structure in cosmology. It does this by consistently including short-scale physics, allowing quantifiable comparisons with real-world data or simulations, and allowing for additional physical processes or modified gravity models.

In this paper, we apply the EFT model developed by \cite{Sailer_2021} (see also \cite{Pourtsidou:2022gsb}) to  intensity mapping surveys, which measure the emission signal of neutral hydrogen in galaxies, in order to trace large-scale structure. Then we combine it with the tree-level bispectrum of 21 intensity mapping, as in \cite{Karagiannis2018,Karagiannis:2019jjx,Karagiannis:2020dpq,Karagiannis_2022}. The survey that we consider is similar to that proposed for HIRAX (see below for details).

The  layout of this paper is as follows. \Cref{sec:hi}  presents an overview of the theoretical models used for the summary statistics considered here, i.e. the power spectrum and the bispectrum.  In \autoref{method} we  present the methodology employed for implementing parameter constraints. Subsequently, the results are presented and elucidated in \autoref{sec:results} and we conclude in \autoref{sec:conc}.

\section{HI intensity power spectrum and bispectrum}
\label{sec:hi}

21cm intensity mapping, also known as neutral hydrogen (HI) intensity mapping (IM), is emerging as a potentially powerful new way to probe the large-scale structure of the post-reionization Universe. By measuring the integrated brightness temperature of the 21cm emission line of  HI trapped in galaxies, this technique evades the need to detect individual galaxies and thus offers a way to rapidly survey huge volumes of the Universe, with exquisite redshift accuracy. A major cosmological survey is in preparation by the Square Kilometre Array Observatory (SKAO) mid-frequency dish array \cite{SKA:2018ckk}. There are complicated observational systematics, in particular major foreground contamination, which are being tackled in initial simulations and observations by precursor experiments (see e.g. \cite{Wang:2020lkn,Cunnington:2022uzo,Cunnington:2023jpq}). 

The SKAO survey will be conducted in single-dish mode, in which the signals from individual dishes are simply added. An alternative is interferometer mode, in which the dish signals are correlated, allowing for the measurement of smaller scales than those accessed by single-dish mode. The HIRAX experiment is being constructed for cosmological surveys in interferometer mode (see \cite{Crichton:2021hlc}) and in this paper we will consider a survey that is similar to that planned for HIRAX. Further details are given in \autoref{table:survey_specs_ITF}. 

\subsection{HI IM bias} \label{sec:halo_bias}

We follow the bias prescription and models in \cite{Karagiannis2018,Karagiannis:2019jjx,Karagiannis:2020dpq,Karagiannis_2022}. The HI IM clustering bias describes the relation between its distribution and that of the dark matter. The $i$-th order bias is given by:
  \begin{equation}\label{eq:bHI}
      b^{\rm HI}_i(z)=\frac{1}{\rho_{\rm HI}(z)}\int_0^\infty  {\rm d} M\, n_{\rm h}(M,z) b_i^{\rm h}(M,z)M_{\rm HI}(M,z),
  \end{equation}
where $\rho_{\rm HI}(z)$ is the HI density \cite{Villaescusa2014}, $n_{\rm h}$ and $b_i^{\rm h}$ are the halo mass function and bias respectively, given by the best-fit results of~\cite{Tinker2008,Lazeyras2016}. $M_{\rm HI}$ is the average HI mass inside a halo of total mass $M$ at redshift $z$ and it is given via the halo occupation distribution (HOD) approach \cite{Cooray2002} and the model of \cite{Castorina2016}: 
\begin{equation}\label{eq:HOD}
    M_{\rm HI}(M,z)=C(z)(1-Y_p)\frac{\Omega_{\rm b}}{\Omega_{\rm m}}\,{\rm e}^{-M_{\rm min}(z)/M}\,M^{q(z)}.
  \end{equation}
Details of the HOD parameters are given in \cite{Karagiannis:2019jjx}. 

The HI IM overdensity is then given by
\begin{equation}\label{eq:deltaG}
   \delta_{\rm HI}(\bx,\tau)= b_1(\tau)\delta_{\rm m}(\bx,\tau) +\veps(\bx,\tau)+\frac{b_2(\tau)}{2} \delta_{\rm m}(\bx,\tau)^2 + \frac{b_{s^2}(\tau)}{2}s(\bx,\tau)^2+\veps_{\delta}(\bx,\tau)\,\delta_{\rm m}(\bx,\tau) ,
\end{equation}
where we dropped label `HI' from $b_i$ for brevity. Here $\delta_{\rm m}$ is the matter overdensity and the $\varepsilon$'s are stochastic terms (see \cite{Karagiannis2018} for details).
The first-order and quadratic biases  are computed from \autoref{eq:bHI}. In the redshift range  $0.7\lesssim z \lesssim 2.6$, we find the fitting formulas 
\begin{align}
\label{eq:linear_bias}
  b_1(z) &=1.333~\, + 0.1809\;z + 0.05302\;z^2 -0.0008822\;z^3   \;,  \\
  b_2(z) &= 0.1802 + 0.1711\;z  + 0.02161\;z^2 + 0.03136\;z^3   \;.  
  \label{eq:quadratic_bias}
\end{align}
The  tidal field bias is modelled as $b_{s^2} =-4(b_1 - 1)/7$, following~\cite{Baldauf2012}.
Note that we marginalise over the bias parameters in each redshift bin: the fitting formulas, based on a halo model framework, are used only to set the fiducial values of the bias parameters.

\subsection{HI IM power spectrum} \label{sec:RSDmodel} 

The HI temperature contrast is 
$\delta_{\rm HI}= {\delta T_{\rm HI}}/{\bar T_{\rm HI}}$,
where $T_{\rm HI}$ is the HI brightness temperature and its background value is related to the background HI number density $\Omega_{\rm HI}$ by \cite{Battye2013}
\begin{align}
\bar T_{\rm HI}(z)=188\,h\,\Omega_{\rm HI}(z) \frac{H_0}{H(z)}~ \mu {\rm K}\,.   
\end{align}
The background number density is currently poorly constrained and we  use the model  
$\Omega_{\rm HI}(z)=4\times 10^{-4}(1+z)^{0.6}$, which is consistent with observations \cite{SKA:2018ckk}. We follow the common practice of   not marginalising over $\bar T_{\rm HI}$, assuming that future observations will determine $\Omega_{\rm HI}$. Given that $\bar T_{\rm HI}$ is degenerate with the bias parameters, this degrades the constraints on those parameters and on $A_s$ \cite{Sailer_2021,Pourtsidou:2022gsb}.

The redshift space HI IM power spectrum is given by
\begin{equation}
P_{\rm HI}(\bk,z)= \bar T_{\rm HI}(z)^2\big[ P(\bk,z) +P_{\rm SN}(z) \Big] +P_{\rm N}(\bk,z)\,.\label{eq:PS_HI}
\end{equation}
At linear order $P(\bk,z)=P^{11}(\bk,z)$, where
\begin{align}
&P^{11}(\bk,z)= Z_1(\bk,z)^2 P_{\rm m}^{\rm L}(k,z) \quad \mbox{with}\\
&Z_1(\bk,z) = b_1(z)+f(z) \mu^2\,.
\label{z1}
\end{align}
Here $\mu=\hat{\bm k}\cdot \hat{\bm n}=k_\|/k$ and $\hat{\bm n}$ is the line of sight; $f=-{\rm d}\ln \delta^{\rm L}_{\rm m}/{\rm d}\ln (1+z)$ is the linear growth rate; 
$P_{\rm SN}$ is the shot noise term and  $P_{\rm N}$ is the instrumental noise (see below). The linear matter power spectrum $P_{\rm m}^{\rm L}$ is computed using the numerical Boltzmann code Cosmic Linear Anisotropy Solving System CLASS \cite{Diego_Blas_2011}.

\begin{figure*}[t]
             \begin{center}
\includegraphics[width=0.7\linewidth ]{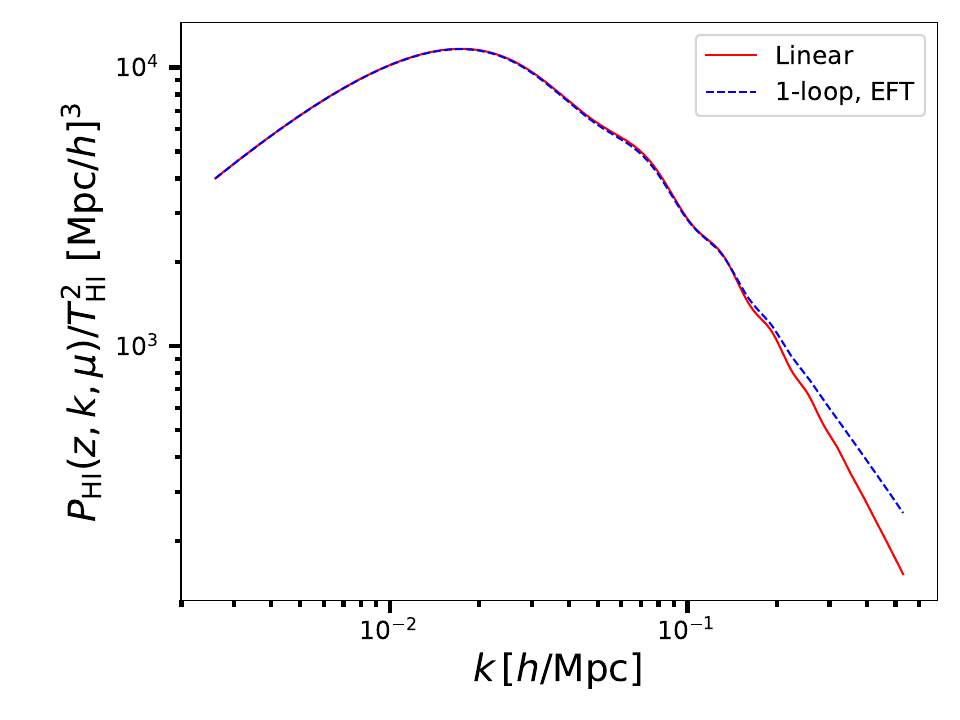}
   \caption{HI IM power spectrum, $P_{\rm HI}/\bar T_{\rm HI}^2$, for tree-level and 1-loop EFT models, at $z=0.8$. }
      \label{fig:EFT}
       \end{center}
    \end{figure*}

In this work we will model the HI IM power spectrum  using SPT at next-to-leading (1-loop) order
\begin{align}
P^{\text{1-loop}}(\bk,z) = P^{11}(\bk,z) + P^{22}(\bk,z) + P^{13}(\bk,z),    
\end{align}
corrected by EFT  counterterms~\cite{Sailer_2021,Pourtsidou:2022gsb}:
\begin{align}
P^{\rm EFT}(\bk,z) = P^{\text{1-loop}}(\bk,z) &+ \big(\alpha_0 + \alpha_2\mu^2 + \alpha_4\mu^4\big)\Big(\frac{k}{k_*} \Big)^2 P^{\rm Zel}_{\rm cb}(\bk,z) \notag \\ 
&+ N_0 + N_2(\mu k)^2 + N_4(\mu k)^4\,,
\label{eq:EFT_powspec}
\end{align}
where we substitute $P(\bk,z)$ with $P^{\rm EFT}(\bk,z)$ in \autoref{eq:PS_HI} and $P_{\rm SN}$ is replaced by the $N_0$ term. Moreover, $\alpha_a$ represent the counter-terms, while $N_2$ and $N_4$ encapsulate small-scale velocities. Following~\cite{Chen:2019lpf}, the Zeldovich approximation of the matter power spectrum to linear order is defined as 
\begin{equation}
P^{\text{Zel}}_{\text{cb}}(\bk,z) = (1 + f\mu^2)^2 P_{\rm m}^{\rm L}(k,z) \,.
\label{eq:zeld_approx}
\end{equation}
The difference between the linear and EFT power spectra is illustrated in \autoref{fig:EFT}. A comprehensive review of the redshift-space counterterms and the free parameters associated with the EFT 1-loop power spectrum is given in \cite{Ivanov:2022mrd}.

\subsection{HI IM bispectrum}

For the HI bispectrum, i.e. the Fourier transform of the three-point function, we use a tree-level model with a phenomenological model of nonlinear RSD, following e.g. \cite{Yankelevich2018,Karagiannis:2019jjx}:
    \begin{align} 
   B_{\rm HI}(\bk_1,\bk_2,\bk_3,z)&= \bar{T}_{\rm HI}(z)^3 \bigg\{ D_\text{fog}^B(\bk_1,\bk_2,\bk_3,z)
   \nonumber \\   &\times\bigg[2Z_1(\bk_1,z)Z_1(\bk_2,z)Z_2(\bk_1,\bk_2,z)P_{\rm m}^{\rm L}(k_1,z)P_{\rm m}^{\rm L}(k_2,z)+2~ \text{perm}\bigg] \nonumber \\&~
   +2P_{\veps\veps_{\delta}}(z)\Big[Z_1(\bk_1,z)P_{\rm m}^{\rm L}(k_1,z)+2~ \text{perm}\Big]+B_{\veps}(z)\bigg\},
   \label{eq:Bgs}
   \end{align}
where the second-order redshift space kernels  is given by:
\begin{align}
Z_2(\bk_i,\bk_j)&=b_1F_2(\bk_i,\bk_j)+f\mu_{ij}^2G_2(\bk_i,\bk_j)+\frac{b_2}{2} +\frac{b_{s^2}}{2}S_2(\bk_i,\bk_j) 
   \nonumber \\ &~
   +\frac{1}{2}f\mu_{ij}k_{ij}\left[\frac{\mu_i}{k_i}Z_1(\bk_j)+\frac{\mu_j}{k_j}Z_1(\bk_i)\right], \label{eq:Z2}
\end{align}
where $\mu_{ij}=(\mu_ik_i + \mu_jk_j)k_{ij}$, $k_{ij}^2=(\bk_i^2+\bk_j^2 )^2$;
$F_2$ and $G_2$ are the SPT kernels \cite{Bernardeau2002}; and the tidal kernel is $S_2(\bk_1,\bk_2)=(\hat{\bk}_1 \cdot\hat{\bk}_2) ^2-1/3$  \cite{McDonald2009,Baldauf2012}. The RSD `fingers of god' damping factor is \cite{Peacock1994, Ballinger1996}
  \begin{align} D_\text{fog}^B(\bk_1,\bk_2,\bk_3,z)&=\exp\Big[-\big(k_1^2\mu_1^2+k_2^2\mu_2^2+k_3^2\mu_3^2\big)\sigma_B(z)^2\Big], \label{eq:fog_BS}
  \end{align}
where $\sigma_B$ characterizes the damping scale, with  fiducial value given by  the linear velocity dispersion $\sigma_\upsilon$. Note that we do not apply an `fog' damping factor to the power spectrum since nonlinear RSD are treated by the EFT counter terms.
Finally, the fiducial values of the stochastic terms in \autoref{eq:Bgs} are \cite{Karagiannis:2019jjx}:
 \begin{equation}\label{eq:poisson_fid}
    P_{\veps\veps_{\delta}}=\frac{b_1}{2\mean{n}_{\rm eff}}, \quad B_{\veps}=\frac{1}{\mean{n}_{\rm eff}^2},
  \end{equation}
where $\mean{n}_{\rm eff}$ is the effective number density and its is given by the HOD formalism used to derive the HI bias, as in \cite{Castorina2016}.

\subsection{Alcock-Paczynski effect}
The Alcock-Paczynski (AP) effect relates the observed angular and redshift separations of objects to the actual cosmological distances, providing a means to investigate the large-scale structure of the Universe. The AP effect is particularly significant in cosmological surveys and clustering analysis, as it allows us to infer information about the cosmological model and geometry of the Universe. To implement this, we need a rescaling factor in the HI power spectrum and bispectrum.

The fiducial cosmology wavevector $\bk_{\rm fid}$ is related to the true wavevector $\bk$ by $\bk_{\rm fid} = \bk_\parallel\alpha_\parallel(z) +\bk_\bot\alpha_\bot(z)$, where  $\alpha_\parallel= H_{\rm fid}(z)/H_{\rm true}(z)$ and $\alpha_\bot$ as $\alpha_\bot =D_{A,\rm true}(z)/D_{A,\rm fid}(z)$. The correlation function, being a dimensionless quantity, is left unchanged by this rescaling \cite{Sailer_2021}. Thus $ P_{\rm HI}^{\rm obs}(\bk_{\rm fid}, z)\, {\rm d}^3\bk_{\rm fid}=  P_{\rm HI}^{\rm true}(\bk, z)\, {\rm d}^3\bk $ leading to:
\begin{align}\label{eq:PS_AP}
P_{\rm HI}^{\rm obs}\left(k, \mu, z\right) &=\left(\frac{H_{\rm true}}{H_{\rm fid}}\right)\left(\frac{D_{A,\rm fid}}{D_{A, \rm true}}\right)^2P_{\rm HI}\left(q,\nu,z \right)  \,.
\end{align}
The fiducial Fourier coordinates $(k,\mu)$ are related to the true values as
\begin{align}
q(k,\mu)&=\beta(\mu)\,k\;, \quad
\nu(k,\mu)=\frac{\mu}{\beta(\mu)}\,\frac{H_{\rm true}}{H_{\rm fid}}\quad \mbox{where}
\label{eq:nu2mu}\\
 \beta(\mu)^2&=\left(\frac{H_{\rm true}}{H_{\rm fid}}\right)^2\mu^2+\left(\frac{D_{A,\rm fid}}{D_{A, \rm true}}\right)^2\left(1-\mu^2\right)\,.  
 \label{eq:nu2mu2}
 \end{align}

Extending \autoref{eq:nu2mu} and \autoref{eq:nu2mu2} to 3 dimensions, we find that \cite{Song2015}:
\begin{align} \label{eq:BS_AP}
B^{\rm obs}_{\rm HI}\left(k_1,k_2,k_3,\mu_1,\phi,z\right)&=\left(\frac{H_{\rm true}}{H_{\rm fid}}\right)^2\left(\frac{D_{A,\rm fid}}{D_{A, \rm true}}\right)^4B_{\rm HI}\left(q_1,q_2,q_3,\nu_1,\phi,z\right). 
\end{align}
This establishes the relationship between the observed and fiducial bispectra, accounting for the AP effect by considering the true and fiducial values of the Hubble parameter $H$ and angular diameter distance $D_A$. Such a formalism is important for making precise measurements and extracting cosmological information from observations of large-scale structure.

\subsection{HI IM window}

The observational window of HI IM surveys is constrained by strong foreground contamination. We apply foreground avoidance filters in the $(k_\perp,k_\|)$ plane \cite{Karagiannis2018,Sailer_2021, Karagiannis_2022}: (a)~a cutoff to excise contaminated long-wavelength radial modes: $k_{\parallel,\,\min} = 0.01\,h/\rm{Mpc}$; and (b)~avoidance of the foreground wedge, where spectrally smooth foregrounds leak into small-scale transverse modes:
\begin{equation}
\label{eq:wedge}
k_{\parallel } \geq A_{\rm wedge}\,k_{\perp} \quad\text{where}\quad  A_{\rm wedge} =\frac{\chi(z)\,H(z)}{c(1+z)}\sin\big[0.61\,N_{\rm w}\,\theta_{\rm b}(z)\big]\,.
\end{equation}
Here $\chi$ is the comoving line-of-sight distance, $N_{\rm w}$ is the number of primary beams away from the beam centre that contaminate the signal, and $\theta_{\rm b}(z)=1.2\lambda_{21}(1+z)/D_{\rm dish}$ is the beam, with $D_{\rm dish}$ the dish diameter. We consider $N_{\rm w} = 1$ \cite{Karagiannis_2022}.

\begin{figure*}[t]
       \begin{center}
\includegraphics[width=0.6\linewidth ]{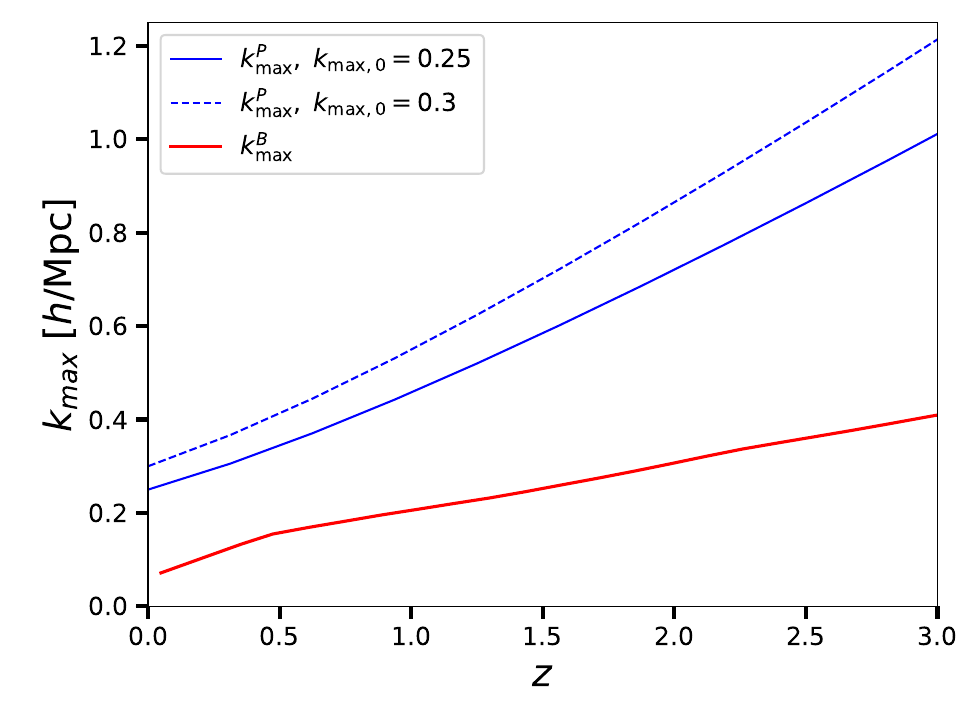}
   \caption{The smallest scale (largest $k$) assumed to be accessible as a function of redshift, using \autoref{eq:eft_kNL} and \autoref{kpmo} for the power spectrum and \autoref{eq:kmax_B} for the bispectrum.}
    \label{fig:eft_kNL}
       \end{center}
    \end{figure*}

The minimum wavenumber, which corresponds to the largest scales probed by a survey of volume $V(z)$ within a redshift bin, is $k_{\rm min}(z)=k_{\rm f}(z)$, where $k_{\rm f}(z)=2\pi /V(z)^{1/3}$ being the fundamental frequency of the survey. The maximum wavenumber (corresponding to the smallest scale) is chosen for the tree-level bispectrum as  \cite{Karagiannis2018} 
\begin{equation}
\label{eq:kmax_B}
    k_{\max}^B(z) =  0.75\,k_{\rm NL}(z)\,, 
\end{equation}
where $k_{\rm NL}$ is the nonlinear scale set by the RMS displacement in the Zel'dovich approximation: 
\begin{equation}
\label{eq:kNL}
    k_{\rm NL}(z)^{-2} =  \frac{1}{6\pi^2}\int_0^{\infty}{\rm d} k\, P^{\rm L}_{\rm m}(k,z).
\end{equation}
The EFT power spectrum probes mildly nonlinear scales with a higher $k_{\max}$. We assume that the maximum scale evolves as
\cite{Baldauf:2016sjb,Chudaykin:2020aoj}  \begin{equation}
\label{eq:eft_kNL}
k_{\rm max}^P(z) = k_{\rm max,0} \, D(z)^{-4/3.3}\,,
\end{equation}
where $D(z)$ is the linear growth factor, normalised to unity at present (i.e. $D(0)=1$). Following \cite{Steele:2021lnz,Ivanov:2022mrd,Chung:2023syw}, we use
\begin{align}\label{kpmo}
k_{\rm max,0}=0.25h~ \mbox{(pessimistic)},~~\,0.3h~ \mbox{(optimistic)} ~~ {\rm Mpc}^{-1}\,. 
\end{align}
The small-scale cuts are shown in \autoref{fig:eft_kNL}.

\section{Fisher forecast formalism}\label{method}

We use the Fisher information matrix \cite{Tegmark_1997, Tegmark_1998} to forecast  precision on model parameters attainable by a HIRAX-like survey. The Fisher matrix for the power spectrum is (e.g. \cite{Karagiannis2018, Karagiannis_2022}) 
\begin{equation}
\label{eq:fisaka}
F_{\alpha\beta}^{P}(z_i)= \frac{1}{2}\sum_{ k}\displaystyle\int_{-1}^{1}{\rm d}\mu \,\frac{\partial P_{\rm HI}^{\rm obs}(\bk,_i)}{\partial \theta_{\alpha}}\frac{\partial  P_{\rm HI}^{\rm obs}(\bk,_i) }{\partial \theta_{\beta}}\frac{1}{\Delta P^2(\bk,_i)}\,
\end{equation}
where the sum is over $k_{\rm min}\leq k \leq k_{\rm max}$. For the bispectrum it is: 
   \begin{equation}\label{eq:fisherBs}
  F_{\alpha\beta}^{B}(z_i)=\frac{1}{4\pi}\!\sum_{k_1,k_2,k_3}\!\int_{-1}^1\!\! {\rm d}\mu_1 \!\! \int_0^{2\pi}\!\! {\rm d} \phi \frac{\partial B_{\rm HI}^{\rm obs}(\bk_1,\bk_2,\bk_3,z_i)}{\partial \theta_{\alpha}}\frac{\partial B_{\rm HI}^{\rm obs}(\bk_1,\bk_2,\bk_3,z_i)}{\partial \theta_{\beta}}\frac{1}{\Delta B^2(\bk_1,\bk_2,\bk_3,z_i)},
  \end{equation}
where the above sum is over all the triangles formed by the wavevectors $k_1, k_2, k_3$, while satisfying $k_{\rm min}\le k_3\le k_2\le k_1 \le k_{\rm max}$. The derivatives are performed over the model parameters $\theta_\alpha$.

We assume that the covariances in the Fisher matrix formalism are Gaussian and diagonal. Then the variances for the two correlators are~\cite{Sefusatti2006,Sefusatti2007}
\begin{align}
\label{eq:sigma_noise}
\Delta P^2(\bk,z) &= \frac{4\pi^2}{V(z)\, k^2 \,\Delta k(z)} \,P_{\text{HI}}^{\text{obs}}(\bk,z)^2 \,,\\
\label{eq:deltaB2}
\Delta B^2(\bk_1,\bk_2,\bk_3,z) &= s_{123}\,\pi\, k_{\text{f}}(z)^3\, \frac{P_{\text{HI}}^{\text{obs}}(\bk_1,z)\,P_{\text{HI}}^{\text{obs}}(\bk_2,z)\,P_{\text{HI}}^{\text{obs}}(\bk_3,z)}{k_1\,k_2\,k_3\,[\Delta k(z)]^3}\,,
\end{align}
where the bin size is chosen as $\Delta k=k_{\rm f}$
and
$s_{123}=6,2,1$ for equilateral, isosceles and non-isosceles triangles,  respectively. In the case of the bispectrum, for degenerate configurations, $k_i=k_j+k_m$, we multiply the variance by a factor of 2~\cite{Chan2017,Desjacques2016}.  Note that higher $k^P_{\rm max,0}$ values than those shown in \autoref{fig:eft_kNL} may still give a reasonably accurate estimate of the power spectrum -- but at the cost of  increasingly non-Gaussian contributions to the power spectrum variance \cite{Chan:2016ehg}. Non-Guassian contributions to the bispectrum occur more readily than in the power spectrum case. For this reason, even with our conservative choice of $k^B_{\rm max,0}$, we 
include higher-order corrections to the variance, as well as taking account of theoretical errors (see \cite{Karagiannis:2019jjx} for a discussion).


\begin{table}[t]
 \centering
 \begin{tabular}{l|c}
Property    & HIRAX\\ \hline\hline
   redshift $z$   & $0.775-2.55$  \\
 $N_{\rm dish}$  & $1024$\\
 $D_{\rm dish}$ [m] & $6$ \\
  $D_{\rm max}$ [km] & $0.25$ \\
 $S_{\rm area}$ [$\rm{deg}^2$]  & $15,000$ \\
 $t_{\rm survey}$ [hrs] & $17,500$ \\
 efficiency $\eta$ & 0.7
 \end{tabular}
 \caption{Specifications for HIRAX interferometer-mode survey (from \cite{Crichton:2021hlc}).}
 \label{table:survey_specs_ITF}
\end{table}

We consider a HI IM survey in interferometer mode, similar to the one proposed for HIRAX~\cite{Newburgh2016,2021SPIE11445E..5OS,2021arXiv210106337K, Crichton:2021hlc} with specifications given in \autoref{table:survey_specs_ITF}.
The instrumental noise term that appears in the power spectrum of \autoref{eq:PS_HI}, is assumed to be Gaussian and it is given in the case of an interferometer by \cite{Zaldarriaga2003b,Tegmark2008,Bull:2014rha}:
\begin{equation}
\label{eq:thermalnoise}
P_{\rm N}(\bm{k}_\perp,z) = T_{\rm sys}(z)^2 \chi(z)^2\lambda(z)\frac{(1+z)}{H(z)}  \left[\frac{\lambda(z)^2}{A_e}\right]^2 \, \frac{S_{\rm area}}{\theta_{\rm b}(z)^2} \,  \frac{1}{ N_{\rm pol} \, n_{\rm b}(\bm{k}_\perp,z) \,t_{\rm survey}} \,,
\end{equation}
where $t_{\rm{survey}}$ is the integration time, $S_{\rm area}$ the sky area, and  $A_{\rm e}$ the effective area, which is determined by the efficiency $\eta$ via $A_{\rm e}=\eta\pi (D_{\rm dish}/2)^2$. 
For the system temperature $T_{\rm sys}$, we add the receiver temperature $T_{\rm rx}=50\,{\rm K}$ to the sky temperature $T_{\rm sky}$, which is obtained from Appendix D of \cite{PUMA2018}.  The baseline distribution $n_{\rm b}(\bm{k}_\perp,z)$ presented in Appendix A of \cite{Karagiannis_2022} is used here.

For the baseline model, we consider the $\Lambda$CDM cosmological parameters, the AP parameters, the growth rate, the bias, the amplitudes of the EFT counter terms and the noise parameters to be free. The counter term and noise parameters are considered nuisance and are marginalised over for each correlator. The final parameter vector then is reduced to
\begin{align}\label{eq:basis}
 \bm{\theta}(z_i)=&\Big\{\Omega_{\rm b},\Omega_{\rm c},h,n_\text{s},A_{\rm s};   
D_A(z_i),H(z_i),f(z_i),b_1(z_i),b_2(z_i), b_{s^2}(z_i)\Big\}\,,
  \end{align}

We also consider the summed information from the power spectrum and bispectrum. In this case we sum the two Fisher matrices, without considering the cross-Fisher between the two correlators, which is assumed to be small, following~\cite{Yankelevich2018}:
\begin{align}
F_{\alpha\beta}^{P+B}(z_i)=F_{\alpha\beta}^{ P}(z_i)+F_{\alpha\beta}^B(z_i)  \,.  
\end{align}
Finally, the total Fisher matrix is a sum over all redshift bins,  assuming that the bins are independent.  Note that during the summation we have to be careful to account for the fact that the redshift-dependent parameters change -- so that they are effectively new parameters for each $z_i$. As a consequence, the dimension of the total Fisher matrix 
\begin{align}
F_{\alpha\beta}^{P+B,{\rm tot}}=\sum_i F_{\alpha\beta}^{P+B}(z_i)  \,,  
\end{align}
is much larger than the dimension of a single-bin matrix. This requires care in the marginalisation.
Further details of the marginalisation and projection in parameter space may be found in \cite{Euclid:2019clj}.

We investigate two further cosmological models that extend $\Lambda$CDM. The constant cosmological parameters for the 3 models are
\begin{align}
\text{$\Lambda$CDM}&:\big\{{\Omega_\text{b}},{\Omega_\text{c}}, h,n_\text{s}, A_{\rm s}  \big\},
\label{eq:LCDM}\\
\text{\wCDM}&:
\big\{{\Omega_\text{b}},{\Omega_\text{c}}, h, n_\text{s}, A_{\rm s}, w_0, w_a \big\}
\quad\mbox{where}~~w_{\rm de}(z)=w_0+w_a\, \frac{z}{1+z}\,,
\label{eq:wcdm}\\
\text{\gCDM}&:\big\{{\Omega_\text{b}},{\Omega_\text{c}}, h, n_\text{s}, A_{\rm s}, \gamma \big\}\quad\mbox{where}~~f(z)=\big[\Omega_{\rm m}(z)\big]^\gamma \,.
\label{eq:gcdm}
\end{align}

In order to find the final constraints on the above parameters we apply  Jacobian transformations to the Fisher matrix that corresponds to the parameter space in \autoref{eq:basis}, after marginalising over the remaining bias parameters:
\begin{equation}
\widetilde{F}_{AB}
=\big(J^{\rm T}\,F^{\rm tot}\, J \big)_{AB}
= \sum_{\alpha,\beta}J_{A\alpha}\, F_{\alpha\beta}^{\rm tot}\,J_{\beta B} \qquad\mbox{where}\quad J_{A\alpha}={\partial \theta_\alpha \over \partial \tilde\theta_A} \,.
\label{eq:Jacobian}
\end{equation}
This will provide the final parameter constraints, from each correlator and their summed signal, in the form $\sigma(\theta_A) = \big[\big(\widetilde{F}^{-1}\big)_{AA}\big]^{1/2}$
.

The fiducial values used for the cosmological parameters are drawn from \Planck cosmic microwave background (CMB) measurements~\cite{Planck2018_cosmo}:  $\Omega_\text{b}h^2 = 0.02237$, $\Omega_\text{c}h^2 = 0.12$, $h = 0.6736$, $n_\text{s} = 0.9649$, and $10^9 A_{\rm s} = 2.1$. 
Fiducial values for the extended model parameters are 
$\gamma = 0.55$, $w_0 = -1$, and $w_a = 0$.

%


\begin{figure*}[t]
\centering
\resizebox{0.8\textwidth}{!}{\includegraphics{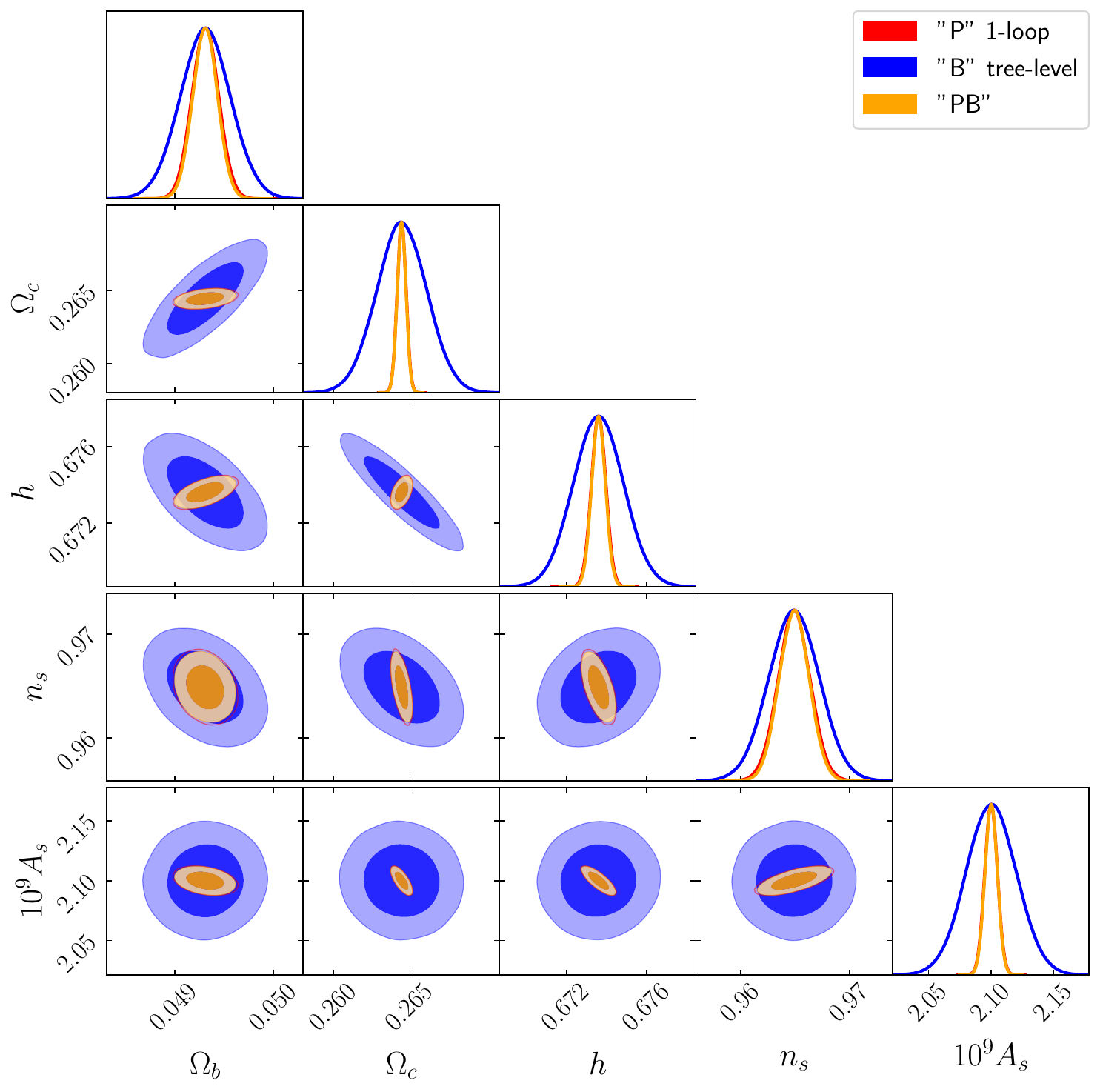}}
\caption{Constraints on $\Lambda$CDM cosmological parameters from the combination of the EFT power spectrum and tree-level bispectrum of a HIRAX-like survey for  $k_{\max,0}=0.25h/$Mpc in \autoref{eq:eft_kNL}. All three cases presented here consider Planck priors (see \autoref{method}) and utilise the signal from the whole redshift range.}
     \label{fig:lcdm_PB_nl}
\end{figure*}

The Fisher matrix results are combined with the data on cosmological parameters obtained from the \Planck   satellite \cite{Planck2018_cosmo}. For this, we use the Markov chain that samples the posterior distribution, obtained from the \Planck webpage\footnote{\url{http://pla.esac.esa.int/pla/\#cosmology}}. The covariance matrix is computed from the chains, specifically for the subset of parameters being analyzed. This matrix is then inverted to obtain the \Planck 2018 Fisher matrix. The Fisher matrices of the power spectrum and bispectrum and their combination are then summed with the latter. The \Planck likelihood is treated as a multivariate Gaussian, appropriate for the free cosmological parameters being studied. 

The cosmological parameter results from  \Planck data include various cosmological models and provide results from MCMC exploration chains, best fits, and tables. Specifically, we use the $\Lambda$CDM chains with baseline likelihoods based on {\sf plikHM\_TTTEEE\_lowl\_lowE},  which we sample using {\sf Getdist}\footnote{\url{https://getdist.readthedocs.io/en/latest/}}.

\section{Results}\label{sec:results}

Here we present and analyse the  results derived from the Fisher matrices for the three cosmological models. 

 \begin{figure*}[t]
 \centering
\resizebox{0.8\textwidth}{!}{\includegraphics{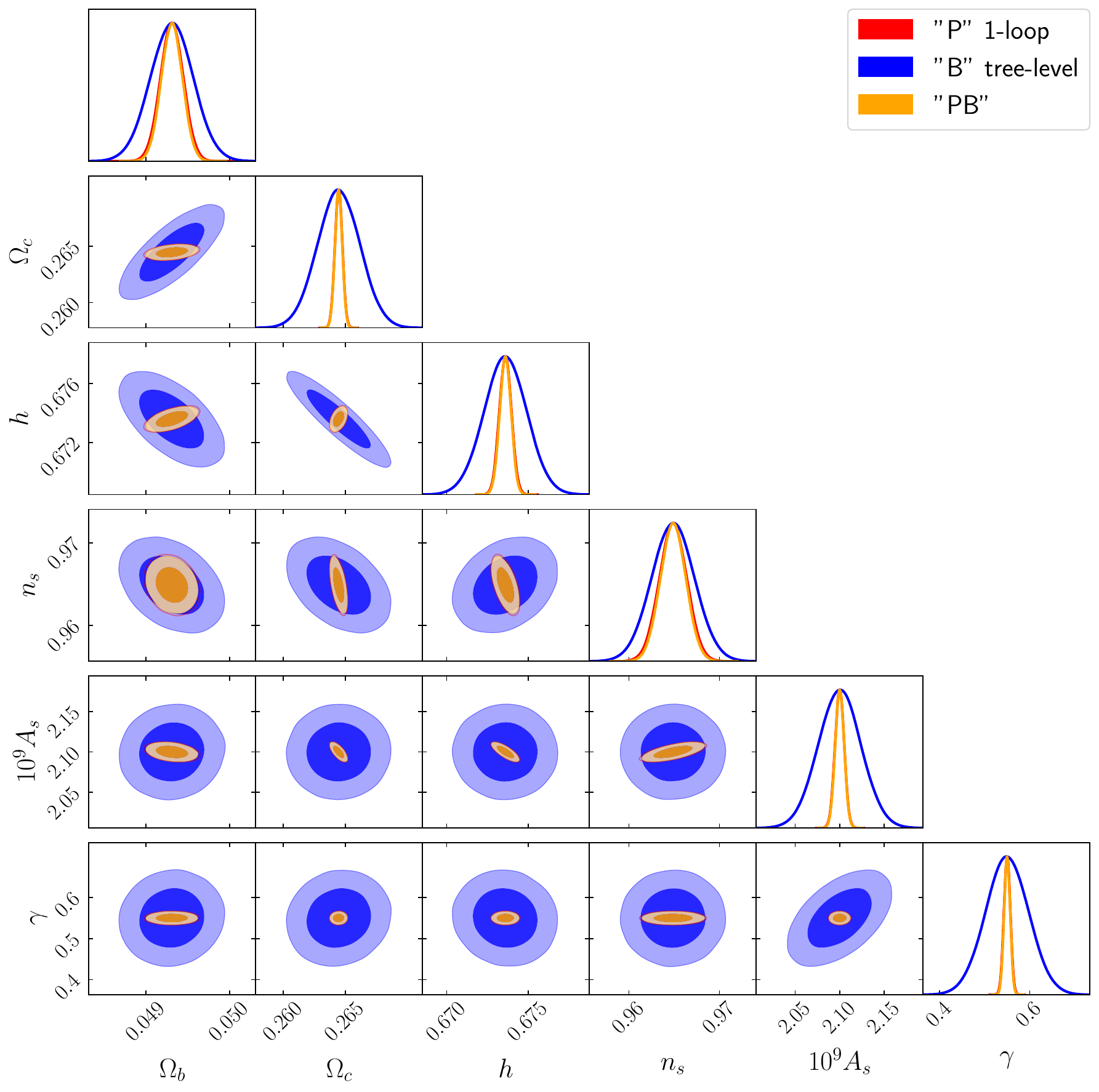}}
\caption{As in \autoref{fig:lcdm_PB_nl}, for the \gCDM model. }
    \label{fig:gammacdm_PB_nl}
    \end{figure*}

\subsection{$\Lambda$CDM}\label{sec:cosmo_res}

We expect the bispectrum to provide  complementary information to the power spectrum. For $\Lambda$CDM, \autoref{fig:lcdm_PB_nl} shows that
bispectrum constraints on cosmological parameters are weaker than from the EFT power spectrum.  
%
Nevertheless there are advantages from combining the 1-loop power spectrum and tree-level bispectrum as shown in \autoref{table:results_table_nl_pessimistic}. This combination enhances parameter constraints for the cosmological parameters $\Omega_b$, $h$ and $n_s$. The constraints achieved with the combined analysis are notably more stringent than those derived from \Planck 2018 CMB data.

The 1-loop power spectrum  and the tree-level bispectrum have different dependencies on the bias and cosmological parameters. This indicates the possibility to break degeneracies and improve constraints, which we intend to explore in follow-up work.  The significant advantages in combining the two correlators is not evident here, due to the relatively small values considered for $k_{\max}^B(z)$ in the tree-level bispectrum analysis, compared to those used for the power spectrum EFT. 




\subsection{Modified gravity}\label{sec:gamma}

The \gCDM model provides a simple test of gravity: the fiducial value $\gamma=0.55$ corresponds to the $\Lambda$CDM and is a good approximation for simple non-clustering models of dark energy in general relativity. This means that a value of $\gamma$ that deviates significantly from 0.55 indicates either a breakdown in general relativity on large scales, or a more complicated dark energy within the Friedmann-Lema\^itre-Robertson-Walker model (see e.g. \cite{Ishak:2018his}).

The 1-loop EFT power spectrum gives reasonable constraints for all considered cosmological parameters, while the bispectrum provides notably less stringent constraints for cosmological parameters and $\gamma$. 


Constraints on the cosmological parameters in the \gCDM model, are naturally weakened, compared to $\Lambda$CDM, by the presence of a new free parameter which affects both the  power spectrum and bispectrum. The EFT power spectrum provides  stronger constraints on the growth rate $f(z)=\Omega_{\rm m}(z)^\gamma$ compared to the  bispectrum, as we show below in \autoref{fig:distances_f_biases}.  However the $f(z)$ constraints do not simply translate to $\gamma$ constraints, since $\gamma$ arises in different ways in the two correlators, and since $\Omega_{\rm m}(z)$ involves the cosmological parameters. \autoref{fig:gammacdm_PB_nl}
shows that in fact, the EFT power spectrum gives better results than the tree-level bispectrum in constraining $\gamma$, while their combination delivers a notable improvement as shown in \autoref{table:results_table_nl_pessimistic}.

\autoref{fig:gammacdm_PB_nl} reveals the possibility to break parameter degeneracies, displayed as different orientation directions in the elliptical contours. Pushing the bispectrum analysis into smaller scales would have improved constraints, once the EFT power spectrum and tree-level bispectrum are combined.

\begin{figure*}[t]
 \centering
\resizebox{0.8\textwidth}{!}{\includegraphics{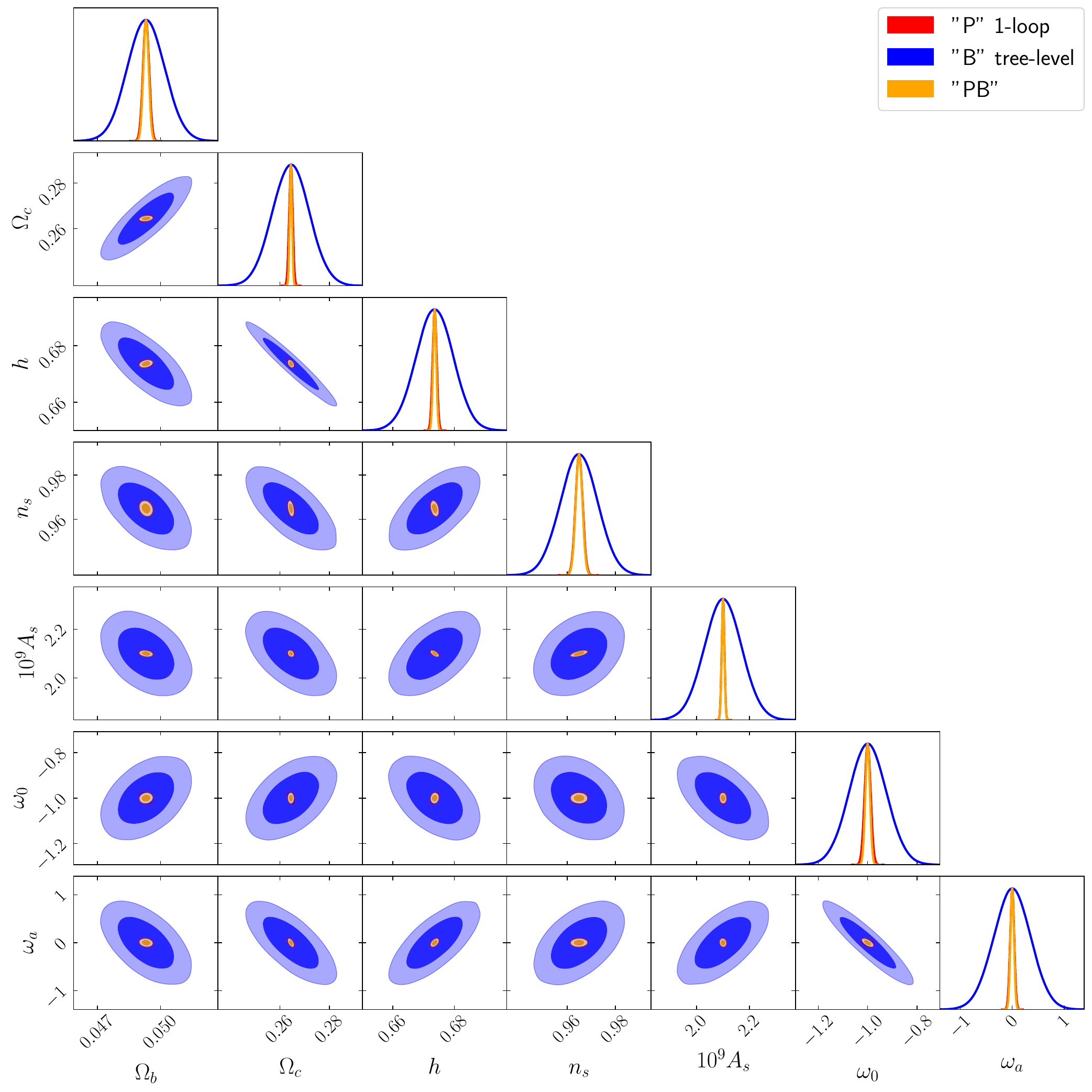} }
\caption{As in \autoref{fig:lcdm_PB_nl}, for the \wCDM model.}
     \label{fig:w0wacdm_P_B_PB_nl}
\end{figure*}

\subsection{Dynamical dark energy}\label{sec:w0wa}

The \wCDM model is a simple model often used to test for deviations from a cosmological constant that are  due to dynamical (or evolving) dark energy (see e.g. \cite{Planck2018_cosmo}). Instead of choosing specific dynamical dark energy models, a two-parameter dark energy equation of state $w(z)$ is used, providing some level of model independence.

With two additional parameters $w_0,w_a$ relative to $\Lambda$CDM, we expect the constraints on cosmological parameters to worsen. This is confirmed by \autoref{fig:w0wacdm_P_B_PB_nl}. It is apparent that bispectrum constraints  degrade even more than those from the power spectrum, where the latter provides more robust constraints to the dark energy parameters than the first. Despite the weakness of the bispectrum constraints, the combination with the power spectrum breaks degeneracies and leads to a noteworthy improvement in constraining $w_0,w_a$, as shown in \autoref{table:results_table_nl_pessimistic}.

\begin{table}[t]
\centerline{
\resizebox{0.5\textwidth}{!}{
\begin{tabular} { l c c c }
\noalign{\vskip 3pt}\hline\noalign{\vskip 1.5pt}\hline\noalign{\vskip 6pt}
 \multicolumn{1}{c}{\bf } & \multicolumn{3}{c}{\bf HIRAX} \\[7pt]
 $[\%]$ &  \bf P & \bf B & \bf P+B \\ 
\noalign{\vskip 3pt}\hline\noalign{\vskip 6pt}
\multicolumn{4}{c}{$\Lambda$CDM}  \\
\noalign{\vskip 3pt}\hline\noalign{\vskip 6pt}
$\sigma(\Omega_{\rm b})/\Omega_{\rm b}$  &  0.29 (0.27) & 0.52  & 0.27 (0.25)\\  
$\sigma(\Omega_{\rm c})/\Omega_{\rm c}$   & 0.12 (0.11) & 0.62  & 0.12 (0.11)  \\  
$\sigma(h)/h$    & 0.06 (0.05) & 0.19 & 0.05 (0.05) \\  
$\sigma(n_{\rm s})/n_{\rm s}$   &   0.16 (0.15)  & 0.24  & 0.15 (0.13)  \\ 
$\sigma(A_{\rm s})/A_{\rm s}$  &  0.25  (0.24) & 0.96  & 0.25 (0.24) \\
\noalign{\vskip 3pt}\hline\noalign{\vskip 6pt}
\multicolumn{4}{c}{\gCDM}  \\
\noalign{\vskip 3pt}\hline\noalign{\vskip 6pt}
$\sigma(\Omega_{\rm b})/\Omega_{\rm b}$  & 0.30 (0.28) &  0.53  & 0.27 (0.26)  \\  
$\sigma(\Omega_{\rm c})/\Omega_{\rm c}$  & 0.12 (0.11) & 0.65 & 0.12 (0.11) \\  
$\sigma(h)/h$   & 0.06 (0.05)  & 0.19  & 0.05 (0.05)\\  
$\sigma(n_{\rm s})/n_{\rm s}$   &  0.17 (0.16) & 0.25 & 0.15 (0.15)\\  
$\sigma(A_{\rm s})/A_{\rm s}$  & 0.26 (0.25) & 1.15  & 0.25 (0.24)   \\
$\sigma(\gamma)/\gamma$  & 1.38 (1.30) & 8.68 & 1.27 (1.22)   \\  
\noalign{\vskip 3pt}\hline\noalign{\vskip 6pt}
\multicolumn{4}{c}{\wCDM}  \\
\noalign{\vskip 3pt}\hline\noalign{\vskip 6pt}
$\sigma(\Omega_{\rm b})/\Omega_{\rm b}$ &  0.32 (0.29)  & 1.79  & 0.27 (0.26)  \\  
$\sigma(\Omega_{\rm c})/\Omega_{\rm c}$  & 0.31 (0.27) & 2.85  &  0.20 (0.18) \\  
$\sigma(h)/h$ & 0.11 (0.10) & 0.91 & 0.09 (0.08) \\  
$\sigma(n_{\rm s})/n_{\rm s}$  & 0.17 (0.15) &  0.80 & 0.15 (0.14)  \\  
$\sigma(A_{\rm s})/A_{\rm s}$   & 0.26 (0.25) & 3.36  & 0.25 (0.24)  \\  
$\sigma(w_0)/|w_0|$   & 1.39 (1.22) & 7.53 & 1.02 (0.93)\\  
$\sigma(w_a)$  & 4.45 (3.91) & 35.69 & 3.80 (3.41)  \\  
\noalign{\vskip 3pt}\hline\noalign{\vskip 1.5pt}\hline\noalign{\vskip 5pt}
\end{tabular}
}
}
\caption{For $\Lambda$CDM, \gCDM and \wCDM: 1$\sigma$ constraints on cosmological parameters and model parameters, from the whole redshift range, using the 1-loop EFT power spectrum (\textbf{P}) and  tree-level bispectrum (\textbf{B}) and their combination (\textbf{P+B}). We use $k_{\max}^B$ as in \autoref{eq:kmax_B} and $k_{\max}^P$ as in \autoref{eq:eft_kNL}, with $k_{\max, 0} = 0.25h$ (pessimistic case). Constraints in brackets are those from the opitimistic case, where $k_{\max, 0} = 0.3h$ is used in \autoref{eq:eft_kNL}.}
    \label{table:results_table_nl_pessimistic}
\end{table}

 \subsection{Summary of results on $\Lambda$CDM, $\gamma$CDM and 
 $w_0w_a$CDM}

\autoref{table:results_table_nl_pessimistic} summarises the 1$\sigma$ constraints on the 3 models, $\Lambda$CDM, \gCDM, and \wCDM. It demonstrates that incorporating the bispectrum enhances the constraints on some parameters compared to using the power spectrum alone. This improvement is mainly evident for the \gCDM and \wCDM model parameters (i.e. $\gamma$, $w_0$ and $w_\alpha$), while for cosmology the gain is minimal. To improve on the latter one would require to extract the bispectrum information from smaller scales, comparable to those used for the power spectrum analysis, by utilising the EFT framework. 

\autoref{table:results_table_nl_pessimistic} shows that there are only small changes  in the power spectrum constraints on the cosmological parameters $\Omega_{\rm b}$, $\Omega_{\rm c}$, $h$, $n_{\rm s}$ and $A_s$ when moving from $\Lambda$CDM to  \gCDM and \wCDM. By contrast, there is a significant increase in error on  $A_s$  (by a factor 1.73) for the bispectrum constraints in \gCDM. In the \wCDM model, the increase in bispectrum errors is even greater, and affects all 5 parameters.
The increase in parameter uncertainties when transitioning from $\Lambda$CDM to \gCDM and \wCDM arises due to the additional correlations introduced among variables as we increase the number of parameters in the model.

\autoref{table:results_table_nl_pessimistic} shows the improvement in the power spectrum constraint when using the optimistic case $k_{\max, 0}=0.3h/$Mpc compared to the pessimistic one (i.e. $k_{\max,0}=0.25/h$Mpc). This is due to the information that resides within the additional small scale modes. However, this improvement is marginal in the case of $\Lambda$CDM and \gCDM, unlike the dark energy model \wCDM, which shows a more notable enhancement in the constraints. This is an indication that the information from the power spectrum saturates, as we venture to smaller scales, due to the thermal noise in the a HIRAX-like instrument.

 \begin{figure*}[t]
       \begin{center}
    \includegraphics[width=1.\linewidth ]{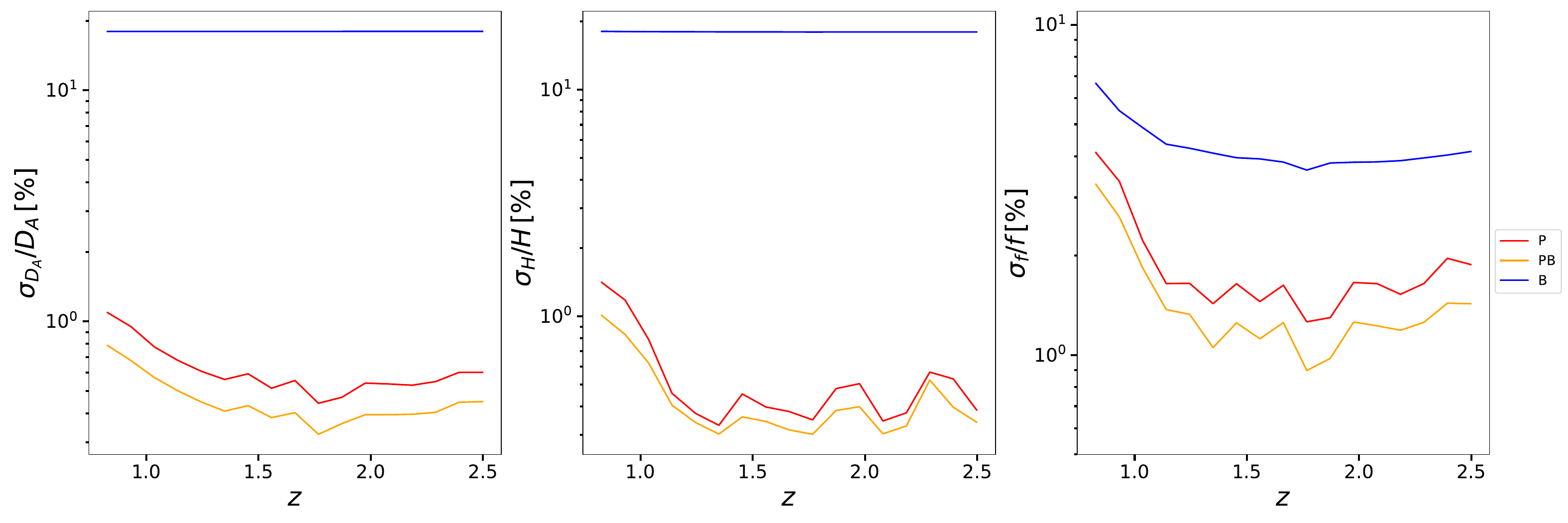}
    \includegraphics[width=1.\linewidth ]{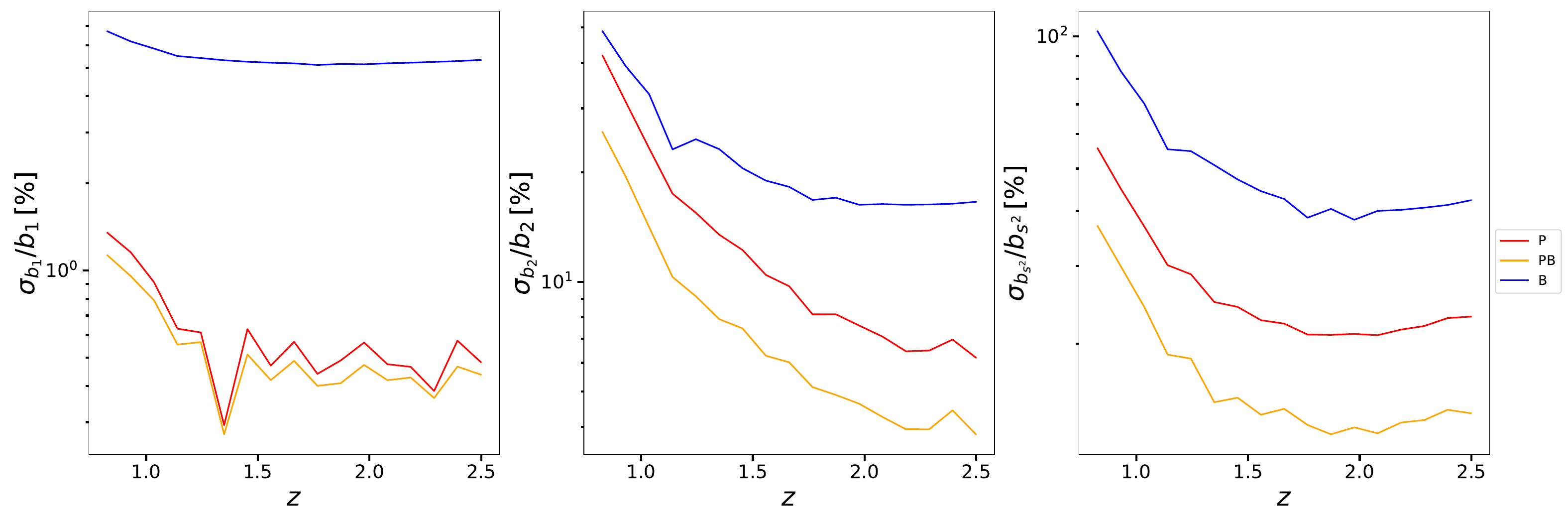}  
\caption{Upper panels: Forecasts of marginalized relative errors on $D_A$, $H$, and $f$ from the power spectrum (red), bispectrum (blue), and their combination (orange), in the pessimistic case (\autoref{kpmo}). Lower panels: Similarly,  for the bias parameters  in \autoref{eq:deltaG}.}
     \label{fig:distances_f_biases}
       \end{center}
    \end{figure*}

 \subsection{Distance and growth rate errors}\label{sec:distance}

In this section, we study the constraints on the angular diameter distance $D_A(z)$ and the Hubble rate $H(z)$ via the AP effect in the standard $\Lambda$CDM model. We also compute errors on the growth rate $f(z)=-{\rm d}\ln D(z)/{\rm d}\ln (1+z)$.

The constraints are derived after marginalising over the nuisance parameters for both correlators, while the remaining model parameters are considered free (\autoref{eq:basis}). \autoref{fig:distances_f_biases} (upper panels) presents the results for the marginalized relative errors as a function of redshift.

The power spectrum delivers remarkably precise measurements, with an uncertainty of less than 1\% across all redshift bins for the angular diameter distance $D_A$ and Hubble parameter $H$, and within 1-2\% for the growth rate $f$.  The bispectrum alone shows a lower precision, yielding relative errors within 10-20\% for $D_A$ and $H$, while for the growth rate the precision is $\sim$5\%, across the redshift range. Consequently, incorporating the bispectrum into the analysis contributes a small improvement in precision on the BAO distance scale parameters and the growth rate.

\subsection{Bias parameters}\label{sec:res_bias}

 Clustering biases affect both the power spectrum and the bispectrum. Here we provide the constraints on biases as a function of redshift before marginalising them to get the cosmology constraints of the previous sections. The results are shown in \autoref{fig:distances_f_biases} (lower panels).


The power spectrum provides a precision of $0.5-1\%$ in most of the redshift range for the HI IM linear bias $b_1$. The joint constraint from the two correlators is at the same level of precision as the power spectrum, indicating the poor performance of the bispectrum. The latter provides a $\sim$5\% precision on $b_1$ constraints, which saturates beyond $z\sim1.5$. For the quadratic bias $ b_2$ and the tidal bias $b_{s^2}$,
the bispectrum provides a precision of $\sim$20\% and $\sim$50\% respectively, where the information saturates beyond $z\sim2$. The power spectrum, on the other hand, exhibits an improvement in the constraints, reaching a precision smaller than $\sim$10\% for $b_2$, at high redshift bins, while for $b_{s^2}$ the relative errors saturate at $\sim$15\% for $z>1.5$. Combining the two correlators offers a significant improvement for these higher order bias parameters, reaching $\sim$4\% and $\sim$5\% at high redshifts for $b_2$ and $b_{s^2}$ respectively. This indicates the great potential offered by a joint EFT power spectrum and bispectrum analysis. 

\section{Conclusions}\label{sec:conc}

This work presents an analysis  to estimate the cosmological precision possible with a HIRAX-like  neutral hydrogen intensity mapping survey, using the 1-loop effective field theory power spectrum and the tree-level bispectrum. In addition to constraints on standard cosmological parameters, we estimate the uncertainties on modified gravity ($\gamma$) and dynamical dark energy ($w_0,w_a$) parameters. Furthermore, we find the errors on the angular diameter distance $D_A(z)$, the Hubble rate $H(z)$ and on the growth rate $f(z)$. Finally, we show the constraints on the linear and higher order clustering biases.

In order to remain within the domain of validity of the tree-level bispectrum and 1-loop EFT power spectrum, we have imposed appropriate scale cuts, as illustrated in \autoref{eq:kmax_B}--\autoref{kpmo}. For the EFT power spectrum, our baseline is a `pessimistic' $k^P_{\rm max}$ cut, with errors also computed for a more optimistic choice.

The 1$\sigma$ marginalised errors of our Fisher forecasts, from the whole redshift range, are summarised in \autoref{table:results_table_nl_pessimistic}. 
For $\Lambda$CDM, the precision from P+B is sub-percent on the cosmological parameters $\Omega_{\rm b}$, $\Omega_{\rm c}$, $h$, $n_{\rm s}$, and $A_{\rm s}$. For the extended models, \gCDM and \wCDM, these constraints naturally degrade, but the precision on the model parameters is high: 1.27\% on the growth index $\gamma$ and (1\%, 3.8\%) on ($w_0$, $w_a$).

Previous work that uses the combination of the power spectrum and bispectrum in cosmological constraints is mainly focused on galaxy surveys (see e.g. \cite{Amendola:2023awr} and references cited therein). In the case of 21cm intensity mapping, there are cosmological forecasts from the 1-loop power spectrum \cite{Sailer_2021,Pourtsidou:2022gsb}, but we are not aware of other work that combines the 1-loop power spectrum with the bispectrum. For the 21cm tree-level power spectrum + bispectrum, the papers \cite{Karagiannis:2019jjx,Karagiannis:2020dpq} investigate future constraints on primordial non-Guassianity.
Compared to the tree-level power spectrum + bispectrum forecasts from our previous work \cite{Karagiannis_2022}, we find that the EFT model improves precision on $\Omega_{\rm b}$, $\Omega_{\rm c}$, and $h$, due to the additional information from small scales, as probed by the 1-loop power spectrum. However, this is not the case for $A_s$ and $n_s$, which are less sensitive to this regime.
The comparison with \cite{Karagiannis_2022} also demonstrates the potential of the EFT power spectrum in breaking parameter degeneracies. This is clearly seen in the contour plots of \autoref{fig:lcdm_PB_nl}--\autoref{fig:w0wacdm_P_B_PB_nl}.

Finally, for the redshift-dependent parameters, we find a sub-percent precision for most redshifts on $D_A, H$, and within 1-2\% for most redshifts on $f$ (\autoref{fig:distances_f_biases}, upper panels). The precision on the linear clustering bias $b_1$ is sub-percent, for most redshifts, while for the quadratic biases it reaches $\sim$5\% from the joint analysis at high redshifts (\autoref{fig:distances_f_biases}, lower panels).

The results presented in this work have relied on certain simplifications that, in a subsequent work, could be relaxed for a more comprehensive analysis. Introducing the 1-loop power spectrum covariance would make the constraints more realistic for future surveys, while extending the reach of the bispectrum, by using the 1-loop Effective Field Theory model (see e.g. \cite{DAmico:2022ukl}), would improve significantly its role in this joint analysis and subsequently the cosmological constraints.

\section*{Acknowledgements}

The authors are supported by the South African Radio Astronomy Observatory
and the National Research Foundation (Grant No. 75415).

\bibliographystyle{elsarticle-num-names}
\bibliography{references}

\end{document}